\documentclass[apj]{emulateapj}
\usepackage{longtable,natbib,graphicx,epstopdf,url,hyphenat}

\newcommand{\nh}{$N_{\rm H}$}    
\newcommand{\ergs}{{\rm erg~s}^{-1}}
\newcommand{\nsa}{NASA-Sloan Atlas}
\newcommand{\cxo}{\textit{Chandra }}

\newcommand{\msun}{M_{\odot}}
\newcommand{\mstar}{M_{\star}}

\newcommand{\gtsima}{$\; \buildrel > \over \sim \;$}
\newcommand{\ltsima}{$\; \buildrel < \over \sim \;$}
\newcommand{\prosima}{$\; \buildrel \propto \over \sim \;$}
\newcommand{\gsim}{\lower.5ex\hbox{\gtsima}}
\newcommand{\lsim}{\lower.5ex\hbox{\ltsima}}
\newcommand{\simgt}{\lower.5ex\hbox{\gtsima}}
\newcommand{\simlt}{\lower.5ex\hbox{\ltsima}}
\newcommand{\simpr}{\lower.5ex\hbox{\prosima}}

\shorttitle{X-ray Selected Black Holes in Dwarf Galaxies}
\shortauthors{Lemons et al.}

\begin{document}

\title{An X-ray Selected Sample of Candidate Black Holes in Dwarf Galaxies}

\author{
Sean~M.~Lemons\altaffilmark{1}
Amy~E.~Reines\altaffilmark{1,3},
Richard~M.~Plotkin\altaffilmark{1},
Elena~Gallo\altaffilmark{1},
Jenny E.~Greene\altaffilmark{2}
}

\altaffiltext{1}{Department of Astronomy, University of Michigan, 1085 South University Ave., Ann Arbor, MI 48109, USA; reines@umich.edu}
\altaffiltext{2}{Department of Astrophysical Sciences, Princeton University, Princeton, NJ 08544, USA}
\altaffiltext{3}{Hubble Fellow}

\begin{abstract}

We present a sample of hard X-ray selected candidate black holes (BHs) in 19 dwarf galaxies.  BH candidates are identified by cross-matching a parent sample of $\sim 44,000$ local dwarf galaxies ($M_{\star} \leq 3 \times 10^9$ M$_\odot$, $z<0.055$) with the {\it Chandra} Source Catalog, and subsequently analyzing the original X-ray data products for matched sources.  Of the 19 dwarf galaxies in our sample, 8 have X-ray detections reported here for the first time.  We find a total of 43 point-like hard X-ray sources with individual luminosities $L_{\rm 2-10keV} \sim 10^{37} - 10^{40}$ erg s$^{-1}$. Hard X-ray luminosities in this range can be attained by stellar-mass X-ray binaries (XRBs), and by massive BHs accreting at low Eddington ratio.  We place an upper limit of $53\%$ (10/19) on the fraction of galaxies in our sample hosting a detectable hard X-ray source consistent with the optical nucleus, although the galaxy center is poorly defined in many of our objects.  We also find that 42\% (8/19) of the galaxies in our sample exhibit statistically significant enhanced hard X-ray emission relative to the expected galaxy-wide contribution from low-mass and high-mass XRBs, based on the $L_{\rm 2-10 keV}^{\rm XRB}-M_\star-$SFR relation defined by more massive and luminous systems.  For the majority of these X-ray enhanced dwarf galaxies, the excess emission is consistent with (but not necessarily due to) a nuclear X-ray source.  Follow-up observations are necessary to distinguish between stellar-mass XRBs and active galactic nuclei powered by more massive BHs.  In any case, our results support the notion that X-ray emitting BHs in low-mass dwarf galaxies may have had an appreciable impact on reionization in the early Universe. 

\end{abstract}


\section{Introduction} 

It is now well-established that massive black holes (BHs) reside in the nuclei of essentially every giant galaxy with a bulge \citep[e.g.,][]{kormendyho2013}, yet the incidence of such BHs in dwarf galaxies is unknown.  Determining the occurrence of massive BHs in present-day low-mass galaxies and studying their properties is currently our best observational probe of the BH seed population in the early Universe \citep{volonteri12,greene12,natarajan14}.  At present, dynamical detections of small BHs in distant dwarf galaxies are not feasible and so we must search for accreting BHs that shine as active galactic nuclei (AGN).  Studying the X-ray radiation from massive BHs in dwarf galaxies is particularly important for models of reionization of hydrogen at very high redshifts \citep[e.g.,][]{volonteri2009} 

Over the last decade, systematic searches using optical spectroscopy from the Sloan Digital Sky Survey (SDSS) have revealed hundreds of low-mass AGN.  The first such studies focused on finding broad-line AGN with virial BH masses $M_{\rm BH} \lesssim 2 \times 10^6~M_\odot$ \citep{greeneho2004,greeneho2007} and narrow-line AGN in galaxies with absolute $g$-band magnitudes $M_g \gtrsim -20$ mag \citep{barthetal2008a}.  More recently, \citet{Reines13} pushed to even lower BH and galaxy masses by searching the spectra of dwarf galaxies with stellar masses $M_{\star} \lesssim 3 \times 10^9 M_\odot$ for both broad and narrow-line signatures of active BHs.  

The optically-selected samples are likely just the tip of the iceberg as they are biased toward relatively high Eddington ratios, and toward galaxies with little ongoing star formation.  High-resolution X-ray and radio observations offer powerful alternatives for discovering massive BHs in low-mass galaxies \citep[e.g.,][]{galloetal2008,galloetal2010,Reines11,reinesdeller2012,milleretal2012,schrammetal2013,reinesetal2014}.

Here we aim to find new massive BH candidates in dwarf galaxies by making use of a wealth of archival data from the {\it Chandra X-ray Observatory}.  With its low background and ability to resolve point sources from diffuse X-ray emission, {\it Chandra} can detect accretion signatures from massive BHs (even nearly quiescent ones) that may not be accessible at other wavebands \citep{soriaetal2006,ghoshetal2008,zhangetal2009,galloetal2010,pellegrini2010}.

This work bears technical similarities to searches for ultraluminous X-ray sources (ULXs) in dwarf galaxies \citep[e.g.,][]{swartz08, kaaret11,prestwich13,plotkin14,brorby14}.  ULXs are {off-nuclear} X-ray sources in excess of the Eddington limit for a $10~M_\odot$ BH ($\sim 10^{39}$ erg s$^{-1}$), the majority of which are now thought to be luminous high mass XRBs.  Luminosities in this regime, however, are also commonly produced by massive BHs accreting at low Eddington ratios \citep{ho09}.

This paper is organized as follows.  Our sample selection is outlined in Section \ref{sec:sample}.  The {\it Chandra} data reduction and photometry is described in Section \ref{sec:data}.  In Section \ref{sec:results}, we present our analysis and results.  A summary of our conclusions and a brief discussion are given in Section \ref{sec:conclusions}.

 \section{Sample Selection}\label{sec:sample}
 
We construct our parent sample of dwarf galaxies using stellar masses provided by the NASA-Sloan Atlas\footnote{http://www.nsatlas.org} (NSA) of local ($z \leq 0.055$) galaxies.  The NSA provides image mosaics, photometry and various galaxy parameters based on a re-analysis of optical and ultraviolet observations from the SDSS and the {\it Galaxy Evolution Explorer (GALEX)}.  Galaxy stellar masses in the NSA are derived from the \texttt{kcorrect} code of \citet{blantonroweis2007}.  Masses are given in units of $M_\odot h^{-2}$ and we adopt $h$ = 0.73. To select dwarf galaxies, we impose a stellar mass upper limit of $M_{\star} \leq 3 \times 10^9 M_\odot$, which is approximately equal to the stellar mass of the Large Magellanic Cloud (LMC).  Our mass threshold is identical to that used by \citet{Reines13} in their optical spectroscopic search for active massive BHs, and leaves us with 44,594 objects.  

After selecting dwarf galaxies from the NSA, we cross-match our parent sample to the list of X-ray sources within the {\it Chandra} Source Catalog \citep[CSC, Release 1.1;][]{Evans10}.  The CSC includes point and compact ($\lesssim 30\arcsec$) sources detected in the Advanced CCD Imaging Spectrometer (ACIS) and High Resolution Camera (HRC) imaging observations from approximately the first eight years of {\it Chandra} operations, {and a source must have a flux at least three times larger than its uncertainty to be included in the CSC \citep[see][]{Evans10}.}  Using a match radius of 5\arcsec\ from the nominal center of the galaxy as given by the NSA, we cull an initial 61 objects from our parent sample {with an X-ray source located close to the optical galaxy center}.

We then examine {the optical properties} of the remaining 61 galaxies in detail to determine if each is indeed a dwarf galaxy.  We check the absolute magnitudes in all SDSS bands given in the NSA and redshifts from the SDSS spectra (and/or NED when available), as well as visually inspect the SDSS images.  This results in the identification of many interlopers: nearby massive galaxies with erroneous mass estimates (these tend to have an obviously incorrect magnitude in an SDSS band), portions of nearby galaxies (including H{\footnotesize II} regions and star clusters), and faint high-redshift galaxies.  Interloping massive galaxies hosting AGN, in particular, will preferentially correlate with X-rays.

For the remaining 31 bona fide dwarf galaxies, we obtain and re-analyze the original X-ray data from the {\it Chandra} archive as described in \S\ref{sec:data} below. {By re-analyzing the X-ray data, and not simply using the CSC catalog products, we are able to tailor the X-ray analysis toward our specific needs. Some of the advantages (which are described in more detail in \S\ref{sec:data}) include the following: we consider  new \textit{Chandra} observations that were taken after the creation of the latest CSC;  we apply the latest \textit{Chandra} calibration files to each image;  we customize source and background photometry apertures to each individual X-ray source; we align the \textit{Chandra} astrometry to the SDSS reference frame when possible, which is important for determining if an X-ray source is consistent with the optical center of each galaxy; and we are able to search for off-nuclear X-ray sources within each galaxy (the initial match to the CSC only isolates X-ray sources 5$\arcsec$ from the optical nucleus).}
 
 To minimize possible contamination from diffuse X-ray emission (e.g., hot gas from star formation) and point-like soft X-ray sources (such as BH and neutron star X-ray binaries in their thermal dominant states, super-soft X-ray sources and/or cataclysmic variables), we exclude sources that are not detected in the hard 2-7 keV band.  As described below, our final sample of dwarfs with hard X-ray selected candidate BHs consists of 19 galaxies and is given in Table~\ref{tab:sample}.

\begin{deluxetable*}{ c c c c c l c c r }
\tablecaption{Dwarf Galaxy Sample }
\tabletypesize{\scriptsize}
\tablehead{
                \colhead{ID}        &
                \colhead{NSAID}        &
                \colhead{SDSS Name}		&
                \colhead{Other Name}        &
                \colhead{$\log \mstar$}        &
                \colhead{$z$}        &
                \colhead{\nh}        &
                \colhead{$M_{g}$}        &
                \colhead{$r_{50}$}        \\             
                \colhead{(1)}        &
                \colhead{(2)}        &
                \colhead{(3)}        &
                \colhead{(4)}        &
                \colhead{(5)}        &
                \colhead{(6)}        &
                \colhead{(7)}        &
                \colhead{(8)}        &
                \colhead{(9)} 	\\ }
 
\startdata
                 1 & 130302 & {J014446.40+170634.0} & MRK 0361 & 9.230 & 0.027 & 5.03 & {$-20.0$} & 2.53 \\
                 2 & 131489 & {J021403.59+275238.0} & NGC 0855 & 9.054 & 0.002 & 6.41 & {$-17.3$} & 16.16 \\
                 3 & 134049 & {J050144.00$-$041718.9} & IC 0399 & 9.343 & 0.013 & 6.46 & {$-19.3$} & 4.91 \\
                 4 & 135806 & {J091721.89+415437.9} & UGC 04904 & 9.068 & 0.006 & 1.03 & {$-17.7$} & 12.41 \\
                 5 & 135954 & {J093401.99+551427.9} & I Zw 18, MRK 0116 & 6.599 & 0.003 & 1.99 & {$-14.9$} & 3.09 \\
                 6 & 34462 & {J103231.89+542403.5} & UGC 05720 & 8.934 & 0.005 & 0.90 & {$-19.2$} & 4.51 \\
                 7 & 36968 & {J103410.14+580349.3} & MRK 1434 & 7.354 & 0.007 & 0.59 & {$-16.5$} & 1.86 \\
                 8 & 90225 & {J105120.73+324558.9} & NGC 3413 & 8.869 & 0.002 & 1.99 & {$-17.6$} & 12.02 \\
                 9 & 3264 & {J115237.19$-$022809.9} & UGC 06850 & 7.812 & 0.003 & 2.25 & {$-16.8$} & 6.39 \\
                 10 & 41753 & {J121326.02+543631.8} & \nodata & 7.607 & 0.008 & 1.43 & {$-15.0$} & 5.16 \\
                 11 & 140999 & {J121539.20+361937.0} & NGC 4214 & 9.071 & 0.001 & 1.49 & {$-19.0$} & 43.08 \\
                 12 & 30830 & {J121923.08+054741.4} & VCC 0344 & 9.417 & 0.007 & 1.58 & {$-17.3$} & 2.44 \\
                 13 & 117784 & {J122111.29+173819.1} & \nodata & 8.850 & 0.007 & 2.68 & {$-17.7$} & 6.84 \\
                 14 & 161603 & {J122219.80+300347.9} & \nodata & 8.565 & 0.002 & 1.73 & {$-15.6$} & 15.78 \\
                 15 & 89394 & {J122548.86+333248.7} & NGC 4395 & 9.102 & 0.001 & 1.35 & {$-18.0$} & 8.00 \\
                 16 & 103196 & {J124516.87+270730.7} & \nodata & 8.928 & 0.004 & 0.71 & {$-18.9$} & 9.51 \\
                 17 & 142389 & {J124911.59+032318.9} & NGC 4701 & 9.241 & 0.002 & 1.91 & {$-17.9$} & 16.51 \\
                 18 & 31028 & {J124957.86+051841.0} & NGC 4713 & 9.199 & 0.002 & 1.95 & {$-18.1$} & 23.16 \\
                 19 & 92811 & {J132501.28+362613.7} & NGC 5143 & 9.342 & 0.019 & 0.92 & {$-18.9$} & 6.11 \\
\enddata 

\tablecomments{\footnotesize Column 1: identification number assigned in this paper.  Column 2: \nsa\  identification number.  Column 3: SDSS name. Column 4: Alternative names.  Column 5: log galaxy stellar mass in units of $\msun$. Column 6: heliocentric redshift. Column 7:  Neutral hydrogen column density in units of $10^{20}$ cm$^{-2}$.  Column 8: absolute $g-$band magnitude corrected for foreground Galactic exctinction. Column 9: Petrosian 50\% light radius in units of arcseconds. With the exception of columns 4 and 7, all values are from the NSA and assume $h = 0.73$.}
 \label{tab:sample}
 
\end{deluxetable*}

\begin{figure*}[!t]
\begin{center}
{\includegraphics[height=8.5in]{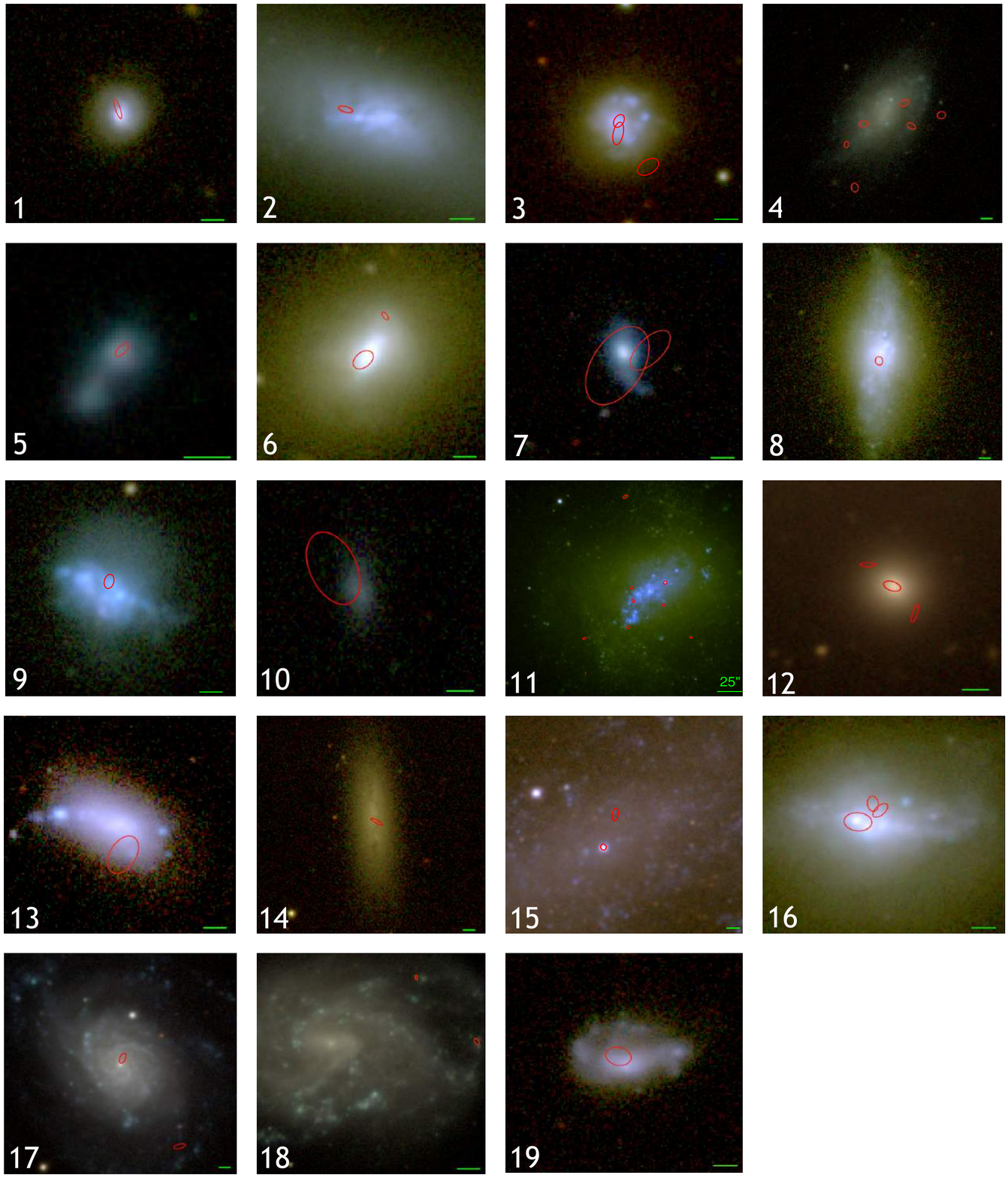}} 
\vspace{-1.5cm}
\end{center}
\caption{\footnotesize SDSS images of our dwarf galaxy sample.  Red ellipses indicate the positions of hard X-ray sources (from {\tt wavdetect}).  Scale bars are 5\arcsec\ in length, except for ID 11 which has a 25\arcsec\ scale bar.
\label{fig:images}}
\end{figure*}

\section{Chandra Data Reduction and Photometry}\label{sec:data}

We obtained observations of 31 dwarf galaxies from the Chandra Data Archive, totaling a cumulative exposure time of {664 ks}, and reduced the data with \texttt{CIAO} version 4.5 \citep{fruscione2006}.  As many of the galaxies were not targeted themselves, their positions are spread over the footprint of the detectors with varying distances from the nominal aim-points. We reprocessed the data to create new level 2 event files and new bad pixel files, taking into account if the observations were taken in FAINT or VFAINT mode. We then checked for background flares and removed time intervals in which the background rate was $> 3\sigma$ above the mean level.  For each observation, we improved the {\it Chandra} astrometry by cross-matching the detected X-ray point sources to the SDSS (which has absolute astrometry accurate to $\lesssim 0\farcs1$), and applied the resulting bore-sight corrections following \cite{zhao05}, as follows.  First, we create X-ray images, including only the 0.3-7.0 keV range, where \cxo is best calibrated. {Then, we run a wavelet detection algorithm on each activated chip using \texttt{CIAO wavdetect} with a sensitivity threshold corresponding to one expected spurious source per field (i.e., for a 1024$\times$1024 pixel image, this corresponds to a \texttt{wavdetect} significance threshold of $\sim$10$^{-6}$).} We cross-match the list {of X-ray sources} with optical sources in the SDSS catalog (with $r < 23$ mag), excluding X-ray sources within three half-light radii of the target galaxy.  If at least two sources were found in common, we generated a new aspect solution file for the {\it Chandra} observation.  This resulted in improved astrometry for $\sim 80\%$ of the observations and the corrections were in the range $\sim 0\farcs01 - 1\farcs01$.  The final astrometric solutions are accurate to between 0\farcs2 and 0\farcs5.  

After registering the \cxo images to the SDSS, we restrict further analysis to the 2-7 keV energy band to search for hard X-ray point sources within three half-light radii of our target galaxies (as the contribution from background X-ray sources is expected to dominate beyond this; see \citealt{swartz08}).   We run \texttt{wavdetect} {on each hard image} with wavelet scales of 1.0, 1.4, and 2.0 pixels, using a 4.5 keV exposure map. {As before, we adjust the \texttt{wavdetect} sensitivity threshold for each image so that we expect at most one  spurious detection per field}. This yielded a total of 60 hard X-ray sources in 26 galaxies. For the positional accuracy of each source centroid, we estimate 95\% error circles ($p_{\rm err}$) using the empirical formula derived by \citet[][see their equation 5]{hong05}; this formula depends on number of hard net counts detected by {\tt wavdetect}, and the location of each source  on the ACIS detector.  The majority of the hard X-ray sources (86\%) have modest positional uncertainties $<$3\arcsec.  The remaining sources are typically located toward an edge of the detector, and can have uncertainties up to $\sim$10\arcsec\ in the most extreme cases.

{We then perform aperture photometry on the hard  X-ray images.}  We adopt circular apertures with radii corresponding to the 90\% encircled energy fraction at 4.5 keV.  The aperture sizes varied from 3\farcs0 - 20\farcs6, depending on the off-axis position of each galaxy.  In most cases, background counts were calculated from an annulus with inner and outer radii of 10\arcsec\  and 30\arcsec\ surrounding the source centroid position (masking out overlapping point-like X-ray sources when present).  For sources farther from the aim-point with larger apertures, we used background annuli with inner and outer radii of 20\arcsec\ and 40\arcsec. {The number of counts in each background annulus are then normalized to the area of each circular source aperture, to estimate the number of expected background counts in each aperture.} We use the 4.5 keV exposure map to estimate effective exposure times at the location of each source on the detector.  The net counts were then divided by 0.9 to correct for the finite aperture size.

{It is possible for {\tt wavdetect} to return a small number of spurious detections, so we further examine each X-ray source.  Any spurious detections would be caused primarily by statistical fluctuations (i.e., the optics and detector characteristics of \textit{Chandra} are well-constrained and accounted for during the reduction procedure).  Since we excluded time intervals subject to unusually high background levels (and limit ourselves to hard X-ray images where the background is typically low; see Table~\ref{tab:xray}), random fluctuations are determined by Poisson statistics.  Using the source and background apertures customized for each X-ray source above, we re-assess which hard X-ray sources are detected at a statistically meaningful level.  To retain a source, we require a number of counts higher than the  background level (within each source aperture) at the $>$95\% confidence level.  For observations with low sky background ($<$0.5 expected counts within the source aperture), we set the detection threshold at 3 counts \citep{gehrels86}.  For sources with large sky backgrounds, we adopt the 95\% confidence level from the Bayesian formalism of \citet{kraft91}, where a typical threshold is 3--4 counts for cases with 0.5--1 expected background counts within the source aperture.}

Sources below the detection threshold {calculated above} are excluded from further analysis. {This resulted in 17 of the 60 detected sources being excluded, and the remaining 43 detected X-ray sources are found in a total of 19 galaxies.  The coordinates of each X-ray source are listed in Table \ref{tab:xray} in order of increasing distance from the nominal optical position of their host galaxy, and we also note in Table~\ref{tab:xray} which sources are also included in the CSC.   Given our adopted 95\% detection confidence limit, we expect 5\% of the sources in Table~\ref{tab:xray} ($\sim$2/43) could still be false positives due to random statistical fluctuations.  Although, 5\% is likely an upper limit, given that many have counts well-above the 95\% confidence limit, all were  deemed detections by \texttt{wavdetect}, and many X-ray sources (30/43) are also included in the CSC.}

\begin{deluxetable*}{ c r r c r c  c c c  c l }
\tablecaption{Hard X-ray Sources}
\tabletypesize{\scriptsize}
\tablehead{
                \colhead{ID}        &
                \colhead{R.A.}        &
                \colhead{Dec}        &
                \colhead{$p_{\rm err}$}        &
                \colhead{Net Counts}        &
                \colhead{N$_{\rm bg}$}     &
                \colhead{Exp. Time}     &
                \colhead{$F_{\rm 2-10 keV}$}        &
                \colhead{$\log L_{\rm 2-10 keV}$}      &
                \colhead{\textit{d}} &
                \colhead{CXO ID}  \\                 
                \colhead{(1)}        &
                \colhead{(2)}        &
                \colhead{(3)}        &
                \colhead{(4)}        &
                \colhead{(5)}        &
                \colhead{(6)}        &
                \colhead{(7)}        &
                \colhead{(8)}    &
                \colhead{(9)}   &
                \colhead{(10)} &
                \colhead{(11)}  \\ }

\startdata
                \textbf{1} & \textbf{26.193495} & \textbf{17.109842} & \textbf{3.20} & $\textbf{3.11}^{+3.63}_{-2.37}$ & \textbf{0.2} & \textbf{13.70} & \textbf{4.93} & \textbf{39.9} & \textbf{1.42} & \textbf{J014446.4+170634} \\
                2 & 33.516998 & 27.877712 & 1.06 & $3.31^{+3.62}_{-2.39}$ & 0.1 & 1.54 & 46.95 & 38.7 & 7.88 & {J021404.0+275239} \\
                 \textbf{3} & \textbf{75.433619} & \textbf{-4.288618} & \textbf{0.40} & $\textbf{21.77}^{+9.37}_{-7.07}$ & \textbf{0.5} & \textbf{33.77} & \textbf{12.80} & \textbf{39.7} & \textbf{0.25} & {\bf J050144.0$-$041718}  \\
                 \nodata & 75.433663 & -4.289359 & 0.40 & $23.99^{+9.75}_{-7.45}$ & 0.5 & \nodata & 14.11 & 39.7 & 2.47 & {J050144.0$-$041721} \\
                 \nodata & 75.431942 & -4.291328 & 0.41 & $18.43^{+8.77}_{-6.45}$ & 0.5 & \nodata & 10.84 & 39.6 & 11.39 & {J050143.6$-$041728} \\
                 4 & 139.337673 & 41.911788 & 0.33 & $79.47^{+16.30}_{-14.07}$ & 0.5 & 51.64 & 33.25 & 39.4 & 7.61 & {J091721.0+415442} \\
                 \nodata & 139.343966 & 41.909299 & 0.34 & $56.13^{+13.98}_{-11.73}$ & 0.5 & \nodata & 23.49 & 39.3 & 11.48 & {J091722.5+415433} \\
                 \nodata & 139.336623 & 41.909045 & 0.31 & $190.58^{+24.32}_{-22.13}$ & 0.5 & \nodata & 79.75 & 39.8 & 11.49 & {J091720.8+415432} \\
                 \nodata & 139.331982 & 41.910328 & 0.79 & $2.80^{+3.64}_{-2.34}$ & 0.5 & \nodata & 1.17 & 38.0 & 22.06 & {\nodata} \\
                 \nodata & 139.346643 & 41.906837 & 0.41 & $15.02^{+8.10}_{-5.76}$ & 0.5 & \nodata & 6.29 & 38.7 & 22.45 & {J091723.2+415424} \\
                 \nodata & 139.345338 & 41.901709 & 0.59 & $3.91^{+3.96}_{-2.88}$ & 0.5 & \nodata & 1.64 & 38.1 & 35.55 & {\nodata} \\
                 \textbf{5} & \textbf{143.508191} & \textbf{55.241235} & \textbf{0.57} & $\textbf{5.51}^{+4.17}_{-3.42}$ & \textbf{0.1} & \textbf{1.28} & \textbf{94.33} & \textbf{39.3} & \textbf{0.50} & \textbf{J093401.9+551428} \\
                 6 & 158.133570 & 54.400612 & 0.40 & $17.54^{+8.60}_{-6.28}$ & 0.2 & 16.38 & 21.08 & 39.2 & 0.81 & {J103232.0+542402} \\
                 \nodata & 158.131360 & 54.403201 & 0.51 & $6.43^{+4.42}_{-3.86}$ & 0.2 & \nodata & 7.73 & 38.8 & 9.79 & {J103231.5+542411} \\
                 \textbf{7} & \textbf{158.542998} & \textbf{58.062962} & \textbf{9.73} & $\textbf{34.70}^{+11.36}_{-9.09}$ & \textbf{2.0} & \textbf{9.47} & \textbf{89.58} & \textbf{40.1} & \textbf{3.02} & \textbf{J103410.1+580346} \\
                 \textbf{\nodata} & \textbf{158.539409} & \textbf{58.063876} & \textbf{9.85} & $\textbf{23.57}^{+9.68}_{-7.38}$ & \textbf{2.0} & \textbf{\nodata} & \textbf{60.84} & \textbf{39.9} & \textbf{5.63} & \textbf{\nodata} \\
                 8 & 162.836691 & 32.765375 & 0.51 & $15.49^{+8.20}_{-5.86}$ & 0.1 & 1.73 & 193.39 & 39.6 & 3.43 & {J105120.8+324555} \\
                 9 & 178.155613 & -2.468589 & 0.53 & $14.39^{+7.97}_{-5.62}$ & 0.1 & 5.06 & 61.57 & 39.3 & 3.81 & {J115237.3$-$022807} \\
                 \textbf{10} & \textbf{183.360607} & \textbf{54.609860} & \textbf{8.95} & $\textbf{91.51}^{+17.37}_{-15.15}$ & \textbf{6.3} & \textbf{29.61} & \textbf{66.81} & \textbf{40.0} & \textbf{6.35} & \textbf{J121326.1+543634} \\
                 11 & 183.920218 & 36.327396 & 0.65 & $8.55^{+4.80}_{-4.69}$ & 0.3 & 26.03 & 7.47 & 37.7 & 17.50 & {J121540.8+361939} \\
                 \nodata & 183.908986 & 36.328876 & 0.41 & $45.28^{+12.73}_{-10.47}$ & 0.3 & \nodata & 39.57 & 38.4 & 17.85 & {J121538.1+361944} \\
                 \nodata & 183.919917 & 36.323634 & 0.76 & $5.38^{+4.18}_{-3.41}$ & 0.2 & \nodata & 4.70 & 37.5 & 19.07 & {\nodata} \\
                 \nodata & 183.909394 & 36.322301 & 0.33 & $235.24^{+26.84}_{-24.65}$ & 0.3 & \nodata & 205.57 & 39.1 & 20.50 & {J121538.2+361921} \\
                 \nodata & 183.921454 & 36.315876 & 0.62 & $6.41^{+4.42}_{-3.86}$ & 0.3 & \nodata & 5.60 & 37.6 & 43.16 & {\nodata} \\
                 \nodata & 183.900242 & 36.312919 & 0.65 & $5.32^{+4.19}_{-3.40}$ & 0.2 & \nodata & 4.65 & 37.5 & 63.39 & {\nodata} \\
                 \nodata & 183.936452 & 36.312645 & 0.52 & $10.86^{+5.11}_{-5.46}$ & 0.3 & \nodata & 9.49 & 37.8 & 81.08 & {J121544.7+361846} \\
                 \nodata & 183.922478 & 36.353828 & 0.70 & $21.89^{+9.39}_{-7.09}$ & 0.3 & \nodata & 19.13 & 38.1 & 101.57 & {J121541.4+362114} \\
                 \nodata & 183.935980 & 36.302550 & 0.80 & $3.10^{+3.63}_{-2.37}$ & 0.2 & \nodata & 2.71 & 37.3 & 106.38 & {\nodata} \\
                 \textbf{12} & \textbf{184.846310} & \textbf{5.794871} & \textbf{0.80} & $\textbf{6.41}^{+4.73}_{-4.17}$ & \textbf{1.4} & \textbf{62.97} & \textbf{2.00} & \textbf{38.3} & \textbf{0.39} & \textbf{J121923.1+054741} \\
                 \nodata & 184.847546 & 5.796062 & 2.00 & $4.20^{+4.27}_{-3.28}$ & 1.4 & \nodata & 1.32 & 38.1 & 6.50 & {\nodata} \\
                 \nodata & 184.845120 & 5.793374 & 1.31 & $3.04^{+3.90}_{-2.75}$ & 1.4 & \nodata & 0.95 & 38.0 & 6.57 & {J121922.8+054735} \\
                 \textbf{13} & \textbf{185.296810} & \textbf{17.636862} & \textbf{11.25} & $\textbf{14.30}^{+7.95}_{-5.60}$ & \textbf{1.3} & \textbf{4.07} & \textbf{69.98} & \textbf{39.9} & \textbf{6.61} & \textbf{J122111.0+173817} \\
                 \textbf{14} & \textbf{185.580669} & \textbf{30.061350} & \textbf{4.76} & $\textbf{3.33}^{+3.62}_{-2.39}$ & \textbf{0.1} & \textbf{4.10} & \textbf{16.16} & \textbf{38.6} & \textbf{3.65} & \textbf{J122219.3+300340} \\
                 \textbf{15} & \textbf{186.453622} & \textbf{33.546937} & \textbf{0.30} & $\textbf{876.17}^{+50.28}_{-48.12}$ & \textbf{0.5} & \textbf{10.58} & \textbf{1882.84} & \textbf{40.1} & \textbf{0.25} & \textbf{J122548.8+333248} \\
                 \nodata & 186.452117 & 33.550481 & 0.77 & $2.83^{+3.64}_{-2.34}$ & 0.5 & \nodata & 6.09 & 37.7 & 13.77 & {\nodata} \\
                 \textbf{16} & \textbf{191.321862} & \textbf{27.125506} & \textbf{0.34} & $\textbf{52.00}^{+13.52}_{-11.27}$ & \textbf{0.2} & \textbf{15.82} & \textbf{68.10} & \textbf{39.4} & \textbf{0.40} & \textbf{J124517.2+270731} \\
                 \nodata & 191.320909 & 27.126584 & 0.40 & $18.66^{+8.82}_{-6.49}$ & 0.2 & \nodata & 24.44 & 39.0 & 4.82 & {\nodata} \\
                 \nodata & 191.320436 & 27.126193 & 0.39 & $20.89^{+9.22}_{-6.91}$ & 0.2 & \nodata & 27.35 & 39.0 & 5.24 & {\nodata} \\
                 17 & 192.298020 & 3.389506 & 0.55 & $5.43^{+4.18}_{-3.41}$ & 0.1 & 9.00 & 1.30 & 38.3 & 3.05 & {J124911.5+032322} \\
                 \nodata & 192.291310 & 3.379165 & 0.73 & $3.22^{+3.63}_{-2.38}$ & 0.1 & \nodata & 7.71 & 38.1 & 42.54 & {\nodata} \\
                 18 & 192.481030 & 5.319315 & 0.98 & $4.42^{+3.92}_{-2.93}$ & 0.1 & 4.40 & 2.23 & 38.5 & 46.22 & {\nodata} \\
                 \nodata & 192.473880 & 5.311602 & 0.55 & $7.72^{+4.60}_{-4.30}$ & 0.1 & \nodata & 3.90 & 38.8 & 61.98 & {J124953.7+051841} \\
                 19 & 201.255760 & 36.437561 & 1.40 & $3.30^{+3.62}_{-2.39}$ & 0.1 & 4.61 & 15.87 & 40.1 & 2.21 & {J132501.3+362615} \\    
             
\enddata 

\tablecomments{\footnotesize Column 1: galaxy ID (see Table 1).  Column 2: right ascension of hard X-ray source in units of degrees.  Column 3: declination of hard X-ray source in units of degrees. Column 4: 95\% positional uncertainty in arcseconds.  The values do not include the 0.2-0.5$\arcsec$ uncertainties from the astrometric correction.  Column 5: net counts in the energy range 2-7 keV, {after applying a 90\% aperture correction to the net counts in each source extraction circle.  Error bars represent  90\% confidence intervals}.  {Column 6: number of background counts expected within each source extraction circle (after applying a 90\% aperture correction).}  Column 7: effective exposure time in kiloseconds at each source position on the detector, determined from 4.5 keV exposure maps.  Column 8: 2-10 keV flux in units of $10^{-15}$ $\ergs~{\rm cm}^{-2}$ corrected for Galactic absorption.  Column 9: log 2-10 keV luminosity in units of $\ergs$ corrected for Galactic absorption.  Column 10: distance from optical center of the galaxy in units of arcseconds. { Column 11: Name of source in CSC catalog, if present. }  Sources are listed in increasing distance from the center of the galaxy, {and sources with positions consistent with the optical nucleus are highlighted in boldface (see \S \ref{sec:spatial})}.}
\label{tab:xray} 
\end{deluxetable*}

Next, we estimate unabsorbed 2-10 keV fluxes using version 4.6b of the Portable, Interactive Multi-Mission Simulator (PIMMS\footnote{See \url{http://heasarc.gsfc.nasa.gov/Tools/w3pimms.html.}}), accounting for Galactic foreground absorption toward each galaxy following \cite{DL90}, and assuming a power law spectrum with photon index $\Gamma$ = 1.8, typical of accreting massive BHs at moderate Eddington ratios \citep{ho09}, hard-state XRBs, as well as many ULXs \citep{swartz08,fengsoria2011}.  Net count rates, unabsorbed fluxes and luminosities are given in Table \ref{tab:xray}.  These values may be considered lower limits as there may be additional absorption intrinsic to the sources. All error bars are quoted at the 90\% confidence level, unless stated otherwise.

Finally, we assess the expected number of foreground/background sources within 3 half-light radii of each galaxy. Following \cite{plotkin14}, we use the published X-ray source counts from the resolved cosmic X-ray background as given by \cite{Mor03}. Starting with the hard (2-10 keV) X-ray flux $S_{dim}$, which corresponds to the dimmest X-ray source in each field, we adopt the expression for the cumulative X-ray flux distribution given in Equation 2 by Moretti et al.\ (2003) to calculate $N(>S)$, i.e. the expected number of X-ray sources per square degree having an X-ray flux brighter than $S$. Finally, for each galaxy field, $N(> S)$ is multiplied by the area enclosed by 3 half-light radii. 
With the exception of one nearby galaxy with a moderately deep exposure (ID 11, NGC 4214; $N_{\rm bk}=4.8$), the expected number of background sources, $N_{\rm bk}$, varies between 0.00025 and 0.74 with a median of 0.016.  Across the sample, we expect a total of 1.4 background contaminants (excluding ID 11).

\section{Analysis and Results}\label{sec:results}

We detect a total of 43 point-like hard X-ray sources in 19 dwarf galaxies (Figure \ref{fig:images}).  Approximately $42\%$ of our sample of galaxies (8/19) have X-ray detections reported here for the first time - we searched the NED databases to identify any X-ray sources that are already known to be background/foreground objects or otherwise reported in the literature.  Notes on individual objects are in the appendix.  

\subsection{Spatial Distribution of X-ray Sources and Candidate Nuclear Sources}\label{sec:spatial}

\begin{figure}[!t]
\begin{center}
{\includegraphics[width=3.5in]{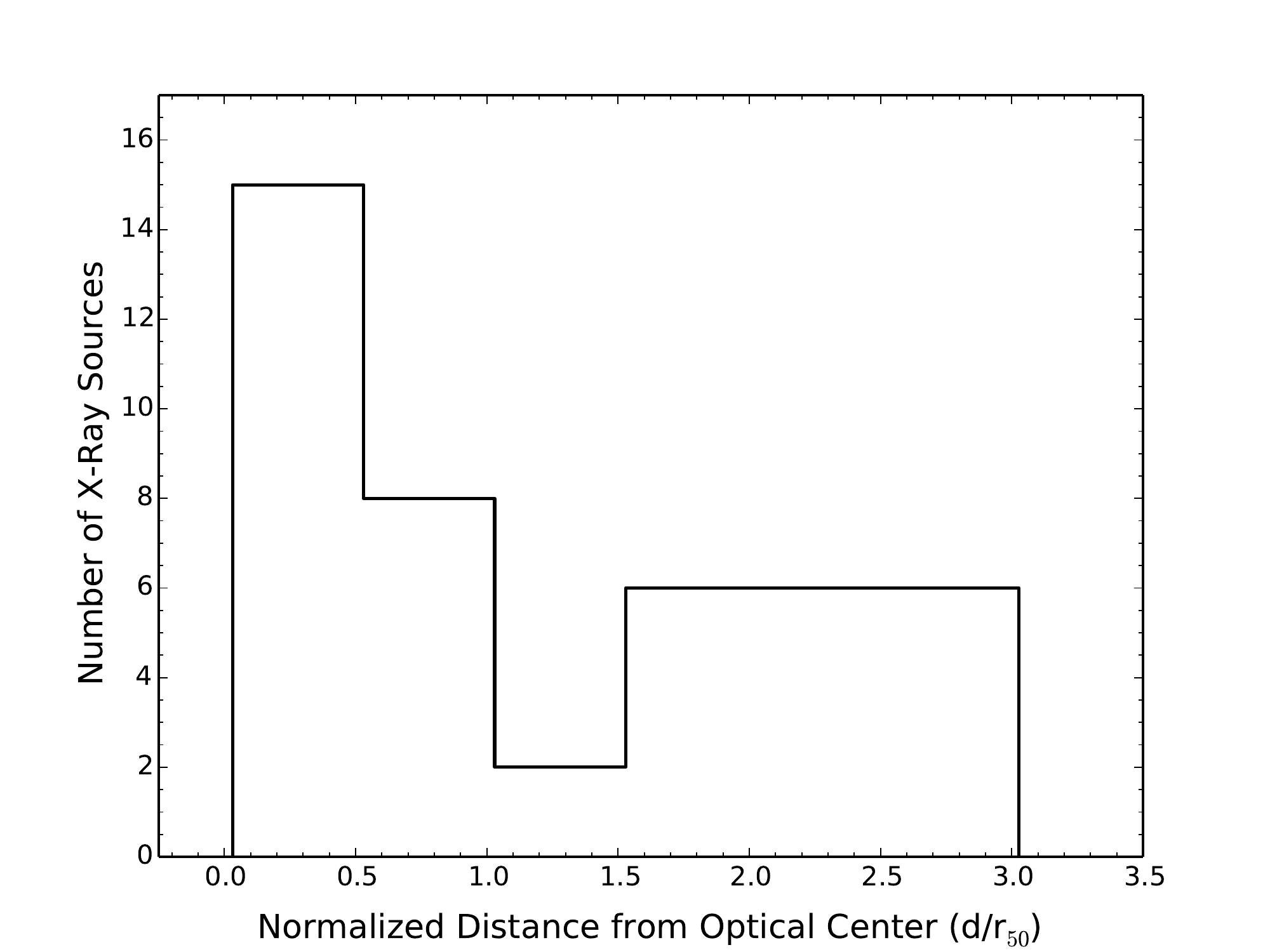}}
\end{center}
\caption{\footnotesize Distribution of the distances of hard X-ray sources from the optical center of their host galaxy, normalized by the half-light radius.    
\label{fig:dist}}
\end{figure}

In Figure \ref{fig:dist}, we show the distribution of the locations of the hard X-ray sources within each galaxy normalized by the half-light radius.  Roughly half of the X-ray sources (23/43) fall within one half-light radius of the optical galaxy center (as given by the NSA).  Such a distribution is expected for the (off-nuclear) luminous XRB population \citep[e.g.,][]{swartz11}.   It is possible that some of the X-ray detections close to the optical center could  be AGN and not XRBs.   However, since the positional accuracy degrades for sources falling toward the edge of the chip, it is difficult to apply a uniform criteria across the entire sample to identify likely nuclear X-ray sources.  Instead, we determine which X-ray sources have 95\% error circles that overlap with the optical center (we add $\sim$0.5\arcsec\ uncertainty from the astrometric correction to the $p_{\rm err}$ estimates), and we find that 11 X-ray sources (in 10 galaxies) are consistent with the nucleus of the host galaxy.  We caution that the optical center of a galaxy may not be well-defined for irregular galaxies or those with star-forming clumps.   

We stress that we consider 11 ``nuclear" X-ray sources to be a generous upper limit, as many of these sources are consistent with the optical center primarily because the error circles on their positions are large (mostly due to galaxies falling toward an edge of the detector).  For example, 5 X-ray sources (in galaxy IDs 7,10,13,14) have positional uncertainties $p_{\rm err} \gtrsim 5$\arcsec\ (with X-ray positions $\sim$3-7\arcsec\ from the optical center), and two of these sources fall within the same galaxy (ID 7).  These X-ray sources could very likely be off-nuclear XRBs.  One previously unidentified X-ray source (in galaxy ID 1; see Appendix) lies $1.4\pm3.2$\arcsec\ from the optical center and could be an intriguing nuclear source.  The remaining sources (IDs 3,5,12,15,16) have much more accurate positions ($p_{\rm err}<1$\arcsec) and lie $<$0.5\arcsec\ from the optical center.  Two of these are newly-identified objects, and three are previously identified X-ray sources including the low-mass Seyfert galaxy NGC 4395 (see Appendix). 

\begin{figure}[!t]
\begin{center}
{\includegraphics[width=3.25in]{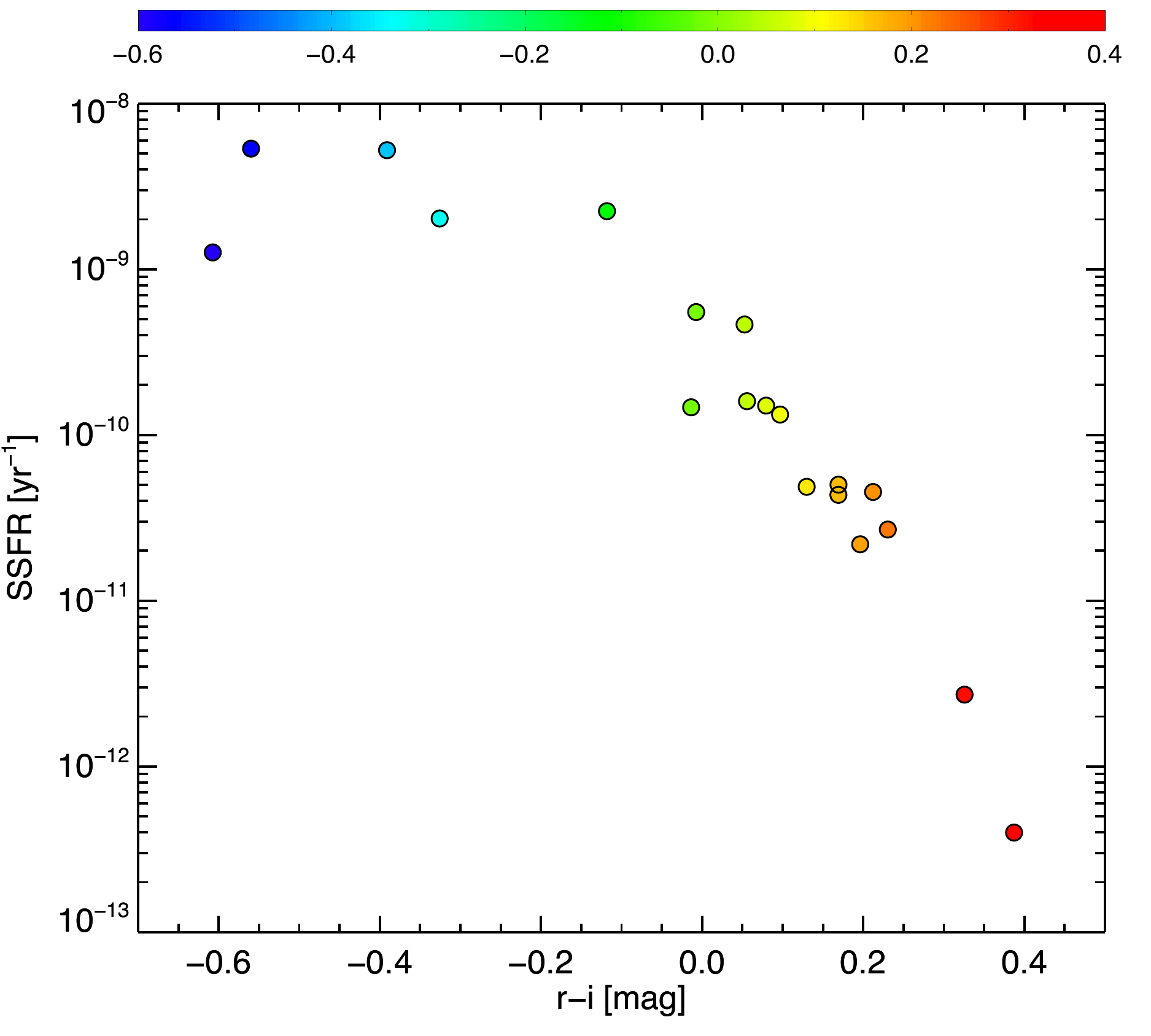}}
\end{center}
\caption{\footnotesize Specific star formation rate versus $r-i$ optical color for the 19 galaxies in our sample.  The color bar indicates $r-i$ color.   
\label{fig:sfr}}
\end{figure}

\subsection{Host Galaxy Properties}\label{sec:host}

\begin{deluxetable*}{c c c r r r c }
\tablecaption{Expected Galaxy-Wide Luminosity from X-ray Binaries}
\tabletypesize{\scriptsize}
\tablehead{
                \colhead{ID}        &
                \colhead{FUV}  &
                 \colhead{$L({\rm FUV)_{obs}}$}  &
                 \colhead{$W_{22}$}   &
                  \colhead{$L(25~\mu$m)}  &
               \colhead{log $SFR$} &
                \colhead{log L$_{\rm 2-10 keV}^{\rm XRB}$}  \\ 
                \colhead{(1)}        &
                \colhead{(2)}        &
                \colhead{(3)}        &
                \colhead{(4)}        &
                \colhead{(5)}     &
                 \colhead{(6)}        &
                \colhead{(7)}  }
\startdata
 1 & 16.44 & 43.45 & 5.01 & 43.17 & $0.58$ & 39.80 \\
2 & 15.20 & 41.67 & 5.39 & 40.74 & $-1.52$ & 38.18 \\
3 & 15.42 & 43.24 & 6.01 & 42.15 & $0.01$ & 39.27 \\
4 & 16.54 & 42.04 & 7.58 & 40.76 & $-1.23$ & 38.30 \\
5 & 15.79 & 41.64 & 7.84 & 39.97 & $-1.67$ & 37.54 \\
6 & 14.24 & 42.83 & 2.53 & 42.66 & $0.04$ & 39.26 \\
7 & 16.42 & 42.34 & 7.45 & 41.07 & $-0.93$ & 38.29 \\
8 & 15.10 & 41.77 & 5.53 & 40.75 & $-1.44$ & 38.10 \\
9 & 15.05 & 42.22 & 4.68 & 41.52 & $-0.88$ & 38.34 \\
10 & 19.57 & 41.13 & $>9.28$ & $<40.40$ & $-2.22$ & 37.13 \\
11 & 11.57 & 42.51 & 2.62 & 41.23 & $-0.76$ & 38.59 \\
12 & 21.14 & 40.37 & $>8.86$ & $<40.42$ & $-2.98$ & 38.38 \\
13 & 16.31 & 42.32 & $>8.07$ & $<40.77$ & $-1.03$ & 38.34 \\
14 & 19.73 & 40.00 & 8.84 & 39.50 & $-3.00$ & 37.54 \\
15 & 13.65 & 41.75 & 5.52 & 40.15 & $-1.56$ & 38.20 \\
16 & 13.48 & 42.89 & 3.96 & 41.84 & $-0.33$ & 38.92 \\
17 & 14.73 & 42.03 & 4.54 & 41.25 & $-1.10$ & 38.46 \\
18 & 14.22 & 42.14 & 5.91 & 40.61 & $-1.16$ & 38.41 \\
19 & 17.37 & 42.79 & 8.12 & 41.64 & $-0.45$ & 38.89
\enddata
 
\tablecomments{\footnotesize  Column 1: galaxy ID (see Table 1). Column 2: {\it GALEX} far-UV AB magnitude.  Column 3: Observed far-UV luminosity
in units of $\ergs$. Column 4: {\it WISE} 22~$\mu$m Vega magnitude.  Column 5: 25~$\mu$m luminosity in units of $\ergs$, using 22~$\mu$m
flux density (in Jy) as a proxy for 25~$\mu$m flux density \citep{Jarrett2013}. Column 6: log SFR in units of $M_\odot~{\rm yr}^{-1}$.
Column 7: Total 2-10 keV luminosity expected from low-mass and high-mass X-ray binaries.  Units are in $\ergs$.}
\label{tab:wise}
 
\end{deluxetable*}

We obtained galaxy stellar masses and half-light radii from the NSA.  The stellar masses span a range of $M_\star \sim 10^{7} - 10^{9.5}~M_\odot$, with a median of $\sim 10^{9}~M_\odot$.  The galaxies in our sample are physically small in addition to being low mass, with half-light radii $\lesssim 2$~kpc.  The galaxies span a range of optical colors and our sample includes spheroids, disks and irregular systems (see Figure \ref{fig:images}).

{We estimate star formation rates (SFRs) of the host galaxies using a combination of far-UV data from {\it GALEX} and 22 $\mu$m data from the {\it Wide-field Infrared Survey Explorer} \citep[{\it WISE};][]{Wright10}.  {\it GALEX} FUV magnitudes were obtained from the NSA and infrared magnitudes were obtained from the {\it WISE} All-Sky Source Catalog (see Table \ref{tab:wise}).  Following \citet{KennicuttEvans12} and \citet{Hao11}, we estimate dust-corrected SFRs as log~SFR = log~$L({\rm FUV)_{corr}} -$ 43.35, where $L{\rm (FUV)_{corr}} = L({\rm FUV)_{obs}} + 3.89~L(25~\mu{\rm m})$ and the luminosities are in units of erg s$^{-1}$.  While the \citet{Hao11} relation employs $25~\mu$m luminosities from the {\it Infrared Astronomical Satellite (IRAS)}, the vast majority of our sample is not detected by {\it IRAS} at $25~\mu$m.  {\it WISE} is sufficiently more sensitive and all but 3 of our sample galaxies have 22 $\mu$m detections.  Given this, and the fact that we expect the flux density ratio between 22~$\mu$m and 25~$\mu$m (in Jy) to be $\sim 1$ for both late-type and early-type galaxies \citep{Jarrett2013}, we use 22~$\mu$m flux densities as a proxy for 25~$\mu$m flux densities.  For the 3 galaxies not detected at 22~$\mu$m in {\it WISE}, we derive SFRs using the observed (uncorrected) FUV luminosities.  The derived SFRs are given in Table \ref{tab:wise}. We plot specific SFRs (SSFR = SFR/$M_\star$) versus $r-i$ optical color for our sample of dwarf galaxies in Figure \ref{fig:sfr}.  As expected, bluer(redder) galaxies have higher(lower) SSFRs. } 

\subsection{Expected Contribution from X-ray Binaries}

\begin{figure*}[!t]
\begin{center}$
\begin{array}{cc}
{\includegraphics[width=3.4in]{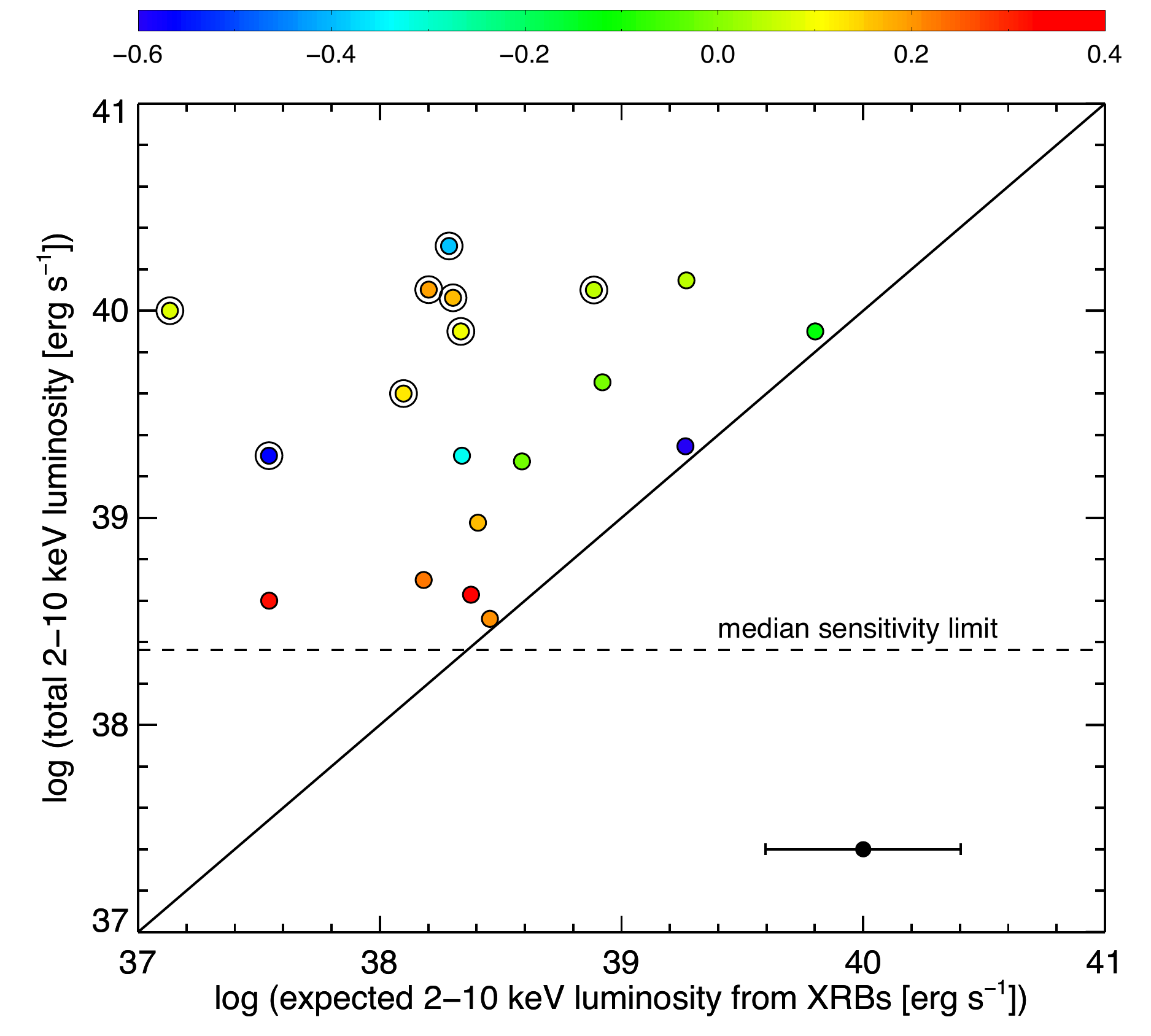}} &
{\includegraphics[width=3.4in]{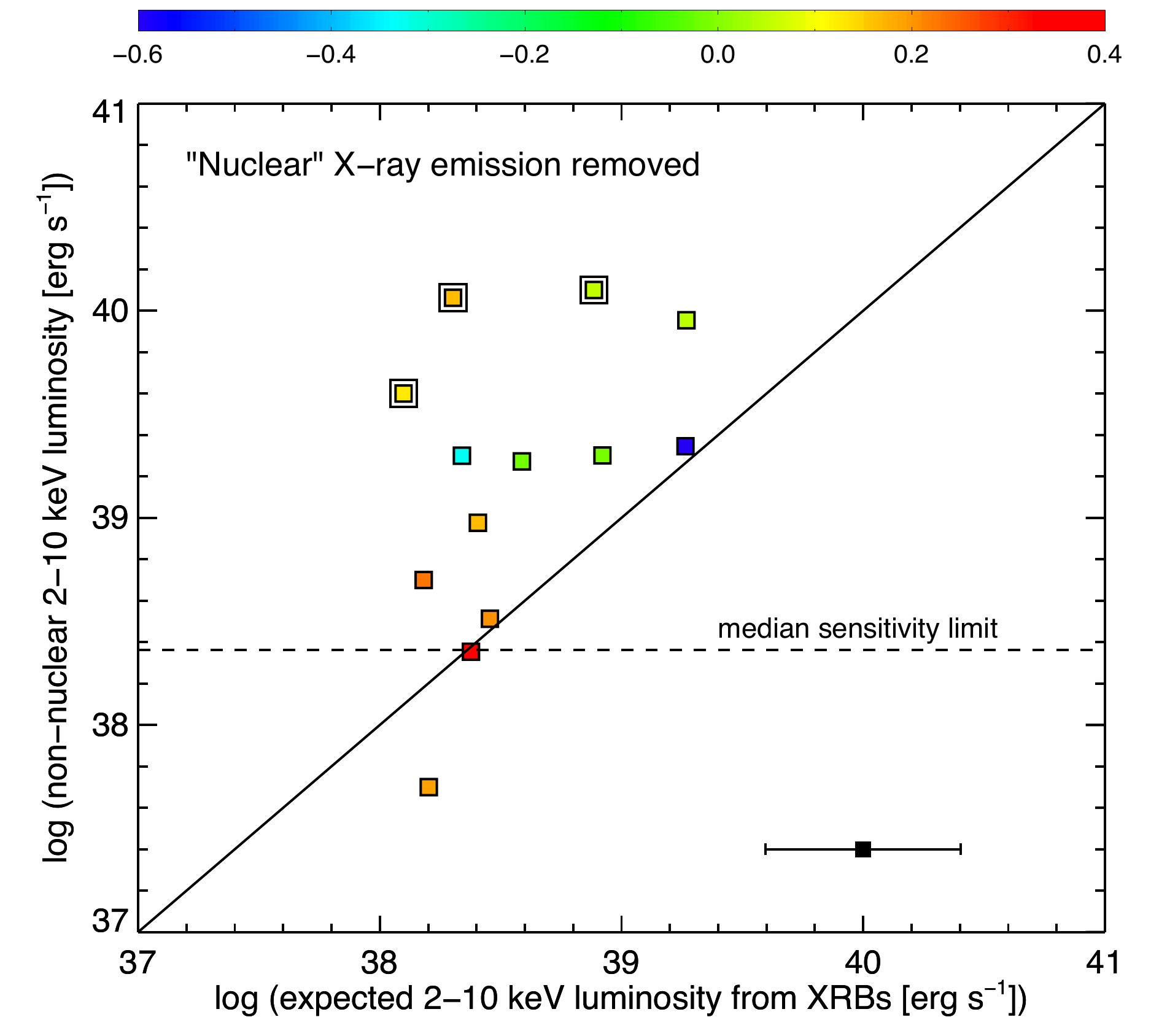}} 
\end{array}$
\end{center}
\caption{\footnotesize 
{\it Left:} Total observed 2-10 keV luminosity versus expected total luminosity from low-mass and high-mass X-ray binaries.  Points are color-coded by $r-i$ color.  The solid line shows the one-to-one relation and the dashed line shows our median X-ray sensitivity limit.  The 1$\sigma$ error bar accounts for systematic uncertainties in stellar masses and SFRs, scatter in the \citet{lehmer10} relation, and errors on $\alpha$ and $\beta$ in the \citet{lehmer10} relation.  Galaxies with statistically significant excess hard X-ray emission ($>3\sigma$, or 1.2 dex) are marked with an additional circle.  
{\it Right:} Same as left panel, except with X-ray emission removed from point sources consistent with the optical center (see \S\ref{sec:spatial}).  
\label{fig:xrb}}
\end{figure*}

The luminosities of individual hard X-ray sources in our sample are in the range $L_{\rm 2-10keV} \sim 10^{37} - 10^{40}$ erg s$^{-1}$. While we are primarily interested in finding candidate {\it massive} BHs, these luminosities could also be attained by stellar mass BHs (or neutron stars) in X-ray binaries (XRBs) with hard X-ray spectra (see, e.g., \citealt{mccr06}).  The X-ray Luminosity Function (XLF) of XRBs in external galaxies \citep{fabbiano06} is known to scale with SFR for late-type galaxies \citep{grimmetal02,grimmetal03,gilfanovetal04,mineo12} and with total stellar mass for early-type galaxies \citep{gilfanov04,humphrey08,lehmer10,lehmer14}, the former being dominated by young, bright high-mass XRBs and the latter by longer-lived low-mass XRBs. 

Given our heterogeneous sample of galaxies, we estimate the expected collective luminosity from XRBs in a given galaxy using stellar masses and SFRs from \S\ref{sec:host} and the relation in \citet{lehmer10}, which accounts for both high-mass and low-mass XRBs:  $L^{\rm XRB}_{\rm 2-10 keV} = \alpha M_\star + \beta$SFR, where $\alpha = (9.05 \pm 0.37) \times 10^{28}$ erg s$^{-1}~M_\odot^{-1}$ and $\beta = (1.62 \pm 0.22) \times 10^{39}$ erg s$^{-1}~(M_\odot~{\rm yr}^{-1})^{-1}$.  While the \citet{lehmer10} relation was derived from more massive and more luminous galaxies, the range of SSFRs in that work is comparable to our sample (SFR/$M_\star \sim 10^{-12} - 10^{-8.5}~{\rm yr}^{-1}$).  In Figure \ref{fig:xrb}, we plot the total observed hard X-ray luminosity versus the expected galaxy-wide luminosity from XRBs. %

We caution that there are large uncertainties associated with the expected total luminosities from XRBs.  The \citet{lehmer10} relation itself has a scatter of $\sim 0.34$ dex.  There are also large uncertainties in the stellar masses and SFRs.  Uncertainties in the stellar masses are expected to be $\sim 0.3$ dex \citep{conroy09} and the scatter in the \citet{Hao11} relation is 0.13 dex.  Stochastic effects are also a concern for galaxies with low stellar masses and/or low SFRs since a small number of luminous XRBs can dominate the total X-ray luminosity.  The $1\sigma$ error bars ($\pm$0.4 dex) in Figure \ref{fig:xrb} account for systematic uncertainties in stellar masses and SFRs, scatter in the \citet{lehmer10} relation, as well as errors on $\alpha$ and $\beta$ in \citet{lehmer10}.    

\subsection{Galaxies with Enhanced X-ray Emission}

{Eight galaxies in our sample (42\%) exhibit excess hard X-ray emission relative to the expected contribution from XRBs (Figure \ref{fig:xrb}).  These galaxies have observed hard X-ray emission $>3\sigma$ (or more than 1.2 dex) higher than expected given their stellar masses and SFRs.  The X-ray sources responsible for the excess emission in five of these cases also have positions consistent with the optical centers of the host galaxies as described in \S \ref{sec:spatial} (IDs 5,7,10,13,15).  ID 15 is the well-studied dwarf Seyfert galaxy NGC 4395 \citep{filippenko89} and ID 5 is the extremely metal-poor galaxy I Zw 18 with a well-studied luminous X-ray source (see Appendix).  To the best of our knowledge, the X-ray sources in galaxies 7, 10 and 13 are reported here for the first time.  However, we note that these galaxies have somewhat irregular morphologies without clearly defined nuclei, and the X-ray positions are not well-constrained.  An additional three galaxies (IDs 4,8,19) exhibit enhanced X-ray emission from non-nuclear sources (see Figure \ref{fig:xrb}).}

\subsection{X-ray Detection Limits}

Given the heterogeneous nature of our galaxy sample, each galaxy has a different luminosity threshold to which an X-ray source could be detected.  To estimate the minimum detectable luminosity, we assume 3 hard X-ray photons (2-7~keV) must be detected within the source aperture (corresponding to a 95\% confidence limit if there is no contribution from the background; \citealt{gehrels86}).  More photons would be required if there is significant contribution from the local background, so we consider the assumption for 3 photons to provide a robust lower limit.  Then, assuming the effective exposure time for each galaxy, the Galactic foreground absorption, the distance to each galaxy, and a power law spectrum with photon index $\Gamma=1.8$, we calculate the minimum 2-10~keV luminosity ($L_{\rm min}$) for which we could detect a hard X-ray source within each galaxy.  The sensitivity limits range from $5\times10^{36}$ to $1\times10^{40}$~$\ergs$, with a median value of  $2.3\times10^{38}$~$\ergs$.
The distribution of observed X-ray luminosities is largely shaped by these detection limits (see Figure \ref{fig:xrb}).
 
\section{Summary and Discussion}\label{sec:conclusions}

We have constructed a sample of 43 hard X-ray selected BH candidates in 19 nearby dwarf galaxies with stellar masses $M_\star \lesssim 3 \times 10^9~M_\odot$.  Eight of the galaxies in our sample have X-ray detections reported here for the first time.  After combining the total hard X-ray flux from all point sources within each galaxy, nearly all of the galaxies have cumulative hard X-ray luminosities that are higher than expected from the combination of low-mass and high-mass XRBs (given each galaxy's stellar mass and SFR).  This high-level of X-ray emission is consistent with  expectations, given the X-ray selected nature of our galaxy sample, and the sensitivity limit of each \textit{Chandra} observation.  For the majority of galaxies, we are typically capable of detecting a hard X-ray source with $L_{\rm 2-10 keV} \gtrsim 10^{38}$~$\ergs$.  Therefore, our study is generally only sensitive to the luminous tail of the XRB population, and more massive accreting BHs. 

Interestingly, even after considering the above bias, we find that a substantial fraction of the galaxies in our sample (42\%) exhibit statistically meaningful enhanced hard X-ray emission, the majority of which have X-ray sources consistent with the optical nucleus.  The dwarf Seyfert 1 galaxy NGC 4395 \citep{filippenko89} is included in this sub-sample, demonstrating that searching for enhanced (cumulative) hard X-ray emission can indeed be a useful diagnostic for identifying AGN in dwarf galaxies.  However, enhanced hard X-ray emission is neither a necessary nor sufficient requirement for an accreting massive BH.  Enhanced galaxy-wide X-ray emission from off-nuclear X-ray sources (thought to be luminous XRBs) has been observed in other studies, especially in low-metallicity dwarfs where the enhancement is up to a factor of 10 larger than for $\sim$solar-metallicity galaxies \citep{prestwich13,brorby14}.  

Follow-up X-ray observations, as well as observations at other wavelengths can help distinguish between stellar-mass XRBs and more massive BHs in our sample.  High-resolution radio observations offer a promising way forward since the ratio of compact radio to hard X-ray emission is significantly higher for accreting massive BHs compared to stellar mass BHs \citep[e.g.,][]{merloni2003}.  The combination of radio and X-ray observations have so far revealed candidate massive BHs in the dwarf galaxies Henize 2-10 \citep{Reines11,reinesdeller2012} and Mrk 709 \citep{reinesetal2014}.  Eleven galaxies in our sample are detected in either the FIRST\footnote{The VLA FIRST Survey: Faint Images of the Radio Sky at Twenty-Centimeters; \citet{becker1995}} or NVSS\footnote{The NRAO VLA Sky Survey, \citet{nvss}} surveys, however the angular resolution is not sufficient to associate radio emission with any particular X-ray source.  Mid-IR diagnostics (e.g., with {\it WISE}) have been used to select luminous AGN with high confidence \citep[e.g.,][]{stern12,assef13}, however using mid-IR colors to identify AGN in dwarf galaxies is not straightforward.  For example, star-forming dwarf galaxies can have mid-IR colors overlapping those of {\it WISE}-selected AGN \citep{izotov14,jarrett11}.
Regardless of the masses of the BHs in our sample, our results have important implications for the impact of X-ray-producing BHs in low-mass galaxies on the epoch of reionization \citep[e.g.,][]{volonteri2009,mirabel2011}.

\acknowledgments

AER, JEG, and EG thank the Kavli Institute for Theoretical Physics for hosting the 2013 workshop ``A Universe of Black Holes", during which this project began.  We thank the anonymous referee for helpful comments and suggestions.  Support for AER was provided by NASA through Hubble Fellowship grant HST-HF2-51347.001-A awarded by the Space Telescope Science Institute, which is operated by the Association of Universities for Research in Astronomy, Inc., for NASA, under contract NAS 5-26555.   This research was supported in part by the National Science Foundation under Grant No.\ NSF PHY11-25915.  This research has made use of data obtained from the Chandra Data Archive and the Chandra Source Catalog, and software provided by the Chandra X-Ray Center (CXC) in the application packages CIAO, ChIPS, and Sherpa.  This publication makes use of data products from the Wide-field Infrared Survey Explorer, which is a joint project of the University of California, Los Angeles, and the Jet Propulsion Laboratory/California Institute of Technology, funded by NASA.  This work has also used observations made with the NASA Galaxy Evolution Explorer. GALEX is operated for NASA by the California Institute of Technology under NASA contract NAS5-98034.  This study has made use of the NASA/IPAC Extragalactic Database (NED) which is operated by the Jet Propulsion Laboratory, California Institute of Technology, under contract with NASA. We are grateful to Michael Blanton and all who helped create the NASA-Sloan Atlas.  Funding for the NASA-Sloan Atlas has been provided by the NASA Astrophysics Data Analysis Program (08-ADP08-0072) and the NSF (AST-1211644).

\appendix

\section{Notes on individual sources}

\noindent
1. Mrk 361 (ObsID 6855; PI Komossa). Belongs the field of view of the Ultra Luminous Infrared Galaxy IIIZw035.  To the best of our knowledge, this is the first paper reporting on the discovery of an X-ray source in the target dwarf galaxy.\\ 

\noindent
2. NGC 855 (ObsID 7095; PI Swartz). Detection of a ULX in this dwarf galaxy was first reported by \cite{swartz08}. \\

\noindent
3. IC 399 (ObsID 9405; PI Gallagher). X-ray sources in this group dwarf galaxy were first reported by \cite{desjardins13}. See also \citealt{tzanavaris14}.\\

\noindent
4. UGC 4904 (ObsID 6729; PI Kulkarni). ToO observation to follow up on a core-collapse supernova in the target galaxy \citep{ofek13}.  Analysis of the point-like X-ray population was first presented by \cite{desjardins14}. \\

\noindent
5. I Zw 18, Mrk 116 (ObsID 805; PI Bomans). A nuclear X-ray source in this metal poor dwarf galaxy was first reported by \cite{ott05}. See also 
\cite{kaaret11, kaaret13,prestwich13,brorby14}. \\

\noindent
6. UGC 5720 (ObsID 9519; PI Mas-Hesse). X-ray sources associated with in this Lyman $\alpha$ emitting galaxy were first reported by \cite{grier11}. See also 
\cite{mineo12, oti12}. \\

\noindent
7. Mrk 1434 (Obsid 3347; PI Barger). Belongs to the field of view of the Lockman Hole-Northwest \citep{yang04}. To the best of our knowledge, this is the first paper reporting on X-ray sources associated the target dwarf galaxy.\\

\noindent
8. NGC 3413 (ObsID 7102; PI Swartz). The X-ray source in this galaxy is part of the \citealt{liu11} catalog (it is not, however, included in \citealt{swartz08}, nor \citealt{swartz09}, as both studies focus on ULXs and thus discard nuclear X-ray sources). \\ 

\noindent
9. UGC 06850 (ObsID 7135; PI Swartz). A bright X-ray source in this group dwarf galaxy was first reported by \cite{swartz09}. See also \citealt{swartz11}.\\ 

\noindent
10. SDSS J121326.02+543631.8 (ObsID 7071; PI Kaaret). Belongs to the field of view of NGC 4194 \citep{kaaret08}.  To the best of our knowledge, this is the first paper reporting on a X-ray source associated with the target dwarf galaxy.\\

\noindent
11. NGC 4214 (ObsID 5197; PI Zezas).  X-ray sources in this nearby dwarf galaxy were first reported by \cite{ghosh06} and \cite{leonidaki10}. See also \cite{liu11} and \cite{mineo12}.\\

\noindent
12. VCC 344 (ObsID 9569; PI Zesaz). Belongs to the field of view of the nearby galaxy NGC 4261 \citep{worrall10} (see also \citealt{gultekin09}). To the best of our knowledge, this is the first paper reporting on X-ray sources in the target dwarf galaxy.\\

\noindent
13. SDSS J122111.29+173819.1 (ObsID 8085; PI Treu). Belongs to the field of view of VCC 437  \citep{galloetal2010}. To the best of our knowledge, this is the first paper reporting on X-ray sources in the target dwarf galaxy.\\

\noindent
14. SDSS J122219.44+300344.1 (ObsID 7853; PI Mathur). Belongs to the field of view of NGC 4308  \citep{milleretal2012}. To the best of our knowledge, this is the first paper reporting on an X-ray source in the target dwarf galaxy.\\

\noindent
15. NGC 4395 (ObsID 5301; PI Moran). This nearby dwarf spiral galaxy harbors the least luminous Seyfert 1 nucleus known.  \cite{oneill} first reported on the X-ray population resolved with \cxo.\\

\noindent
16. SDSS J124517.25+270732.1 (Obsid 13928; PI Reines). Reines et al., in prep. \\

\noindent
17. NGC 4701 (ObsID 7148; PI Soria). \cite{desroches09} first reported on a nuclear X-ray source in this nearby late type dwarf. Here we report on an additional off-nuclear source. See also \cite{liu11}.\\

\noindent
18. NGC 4713 (ObsID 4019; PI Satyapal). \cite{dudik05} first reported on the lack of a nuclear X-ray source in this dwarf. Here we report on off-nuclear sources. See also \cite{liu11}.\\

\noindent
19. NGC 5143 (ObbsID 4055; PI Sambruna). Belongs to the field of view of the FRI 1322+36. To the best of our knowledge, this is the first paper reporting on X-ray sources in the target dwarf galaxy.\\

\end{document}